\documentclass[
  twocolumn,
  prl,
  showpacs,
  amsmath,
  amssymb,
  superscriptaddress,
  nofootinbib,
  floatfix
]{revtex4}

\usepackage{mathtools}
\usepackage{bm}
\usepackage{dsfont}
\usepackage{graphicx}
\usepackage{pifont}
\usepackage{textcomp}
\usepackage{gensymb}
\usepackage{accents}
\usepackage{upgreek}
\usepackage{array}
\usepackage{hhline}
\usepackage{tabularx}
\usepackage{color}

\makeatletter
\def\NAT@bibsetnum#1{%
 \setlength{\topsep}{\z@}%
 \NATx@bibsetnum{#1}%
}%
\renewenvironment{thebibliography}[1]{%
 \NAT@thebibliography{#1}%
 \@clubpenalty\clubpenalty
 \let\@TBN@opr\present@bibnote
 \@FMN@list
}{%
 \@endnotesinbib
 \edef\@currentlabel{\arabic{NAT@ctr}}%
 \NAT@endthebibliography
 \global\let\auto@bib\@empty
}
\newcommand*{\supplementarystart}{%
  \close@column@grid%
  \clearpage%
  \onecolumngrid%
  \setcounter{enumiv}{0} % resets counter for references
  \setcounter{equation}{0} % resets counter for equations
  \setcounter{figure}{0} % resets counter for figs
  \setcounter{table}{0} % resets counter for tables
  \setcounter{page}{1}
  \c@secnumdepth=4
  \renewcommand{\theequation}{s\arabic{equation}} % equations numbered with S...
  \renewcommand{\bibnumfmt}[1]{[s##1]} % bibtems [S...]
  \renewcommand{\@onlinecite}{s\citealp} % citations [S...]
  \renewcommand{\cite}[1]{{[}\onlinecite{##1}{]}}
  \renewcommand{\thefigure}{s\arabic{figure}}
  \renewcommand{\thetable}{s\Roman{table}}
  \renewcommand{\thepage}{s\arabic{page}}
}
\makeatother

\newcommand{\s}{\sum\limits}

\newcommand{\be}{\begin{equation}}
\newcommand{\e}{\end{equation}}
\newcommand{\beml}{\begin{subequations}}
\newcommand{\eml}{\end{subequations}}
\newcommand{\beq}{\begin{eqnarray}}
\newcommand{\eq}{\end{eqnarray}}
\newcommand{\ba}{\begin{array}}
\newcommand{\ea}{\end{array}}
\newcommand{\bpm}{\begin{pmatrix}}
\newcommand{\epm}{\end{pmatrix}}
\newcommand{\bc}{\begin{cases}}
\newcommand{\ec}{\end{cases}}
\newcommand{\lt}{\left}
\newcommand{\rt}{\right}

\newcommand{\ep}{\varepsilon}

\newcommand{\bb}{\boldsymbol}
\newcommand{\h}{^\dagger}
\newcommand{\0}{^\phantom{\dagger}}

\newcommand{\nimp}{n_\textrm{imp}}

\DeclareMathOperator{\tr}{Tr}

\DeclareMathOperator{\im}{Im}
\DeclareMathOperator{\re}{Re}

\begin{document}

\title{L\'evy flights due to anisotropic disorder in graphene}

\author{S.~Gattenl{\"o}hner}
\affiliation{Radboud University, Institute for Molecules and Materials, NL-6525 AJ Nijmegen, The Netherlands}

\author{I.\,V.~Gornyi}
\affiliation{Institut f{\"u}r Nanotechnologie, Karlsruhe Institute of Technology, 76021 Karlsruhe, Germany}
\affiliation{A.\,F.~Ioffe Physico-Technical Institute, 194021 St.\,Petersburg, Russia}
\affiliation{\mbox{Institut f\"ur Theorie der Kondensierten Materie, Karlsruhe Institute of Technology, 76128 Karlsruhe, Germany}}
\affiliation{L.\,D.~Landau Institute for Theoretical Physics RAS, 119334 Moscow, Russia}

\author{P.\,M.~Ostrovsky}
\affiliation{Max Planck Institute for Solid State Research, Heisenbergstr.\,1, 70569 Stuttgart, Germany}
\affiliation{L.\,D.~Landau Institute for Theoretical Physics RAS, 119334 Moscow, Russia}

\author{B.~Trauzettel}
\affiliation{\text{Institut f\"ur Theoretische Physik und Astrophysik, Universit\"at W\"urzburg, 97074 W\"urzburg, Germany}}

\author{A.\,D.~Mirlin}
\affiliation{Institut f{\"u}r Nanotechnologie, Karlsruhe Institute of Technology, 76021 Karlsruhe, Germany}
\affiliation{\mbox{Institut f\"ur Theorie der Kondensierten Materie, Karlsruhe Institute of Technology, 76128 Karlsruhe, Germany}}
\affiliation{Petersburg Nuclear Physics Institute,188300 St.\,Petersburg, Russia.}
\affiliation{L.\,D.~Landau Institute for Theoretical Physics RAS, 119334 Moscow, Russia}

\author{M.~Titov}
\affiliation{Radboud University, Institute for Molecules and Materials, NL-6525 AJ Nijmegen, The Netherlands}

\begin{abstract}
We study transport properties of graphene with anisotropically distributed on-site impurities (adatoms) that are randomly placed on every third line drawn along carbon bonds. We show that stripe states characterized by strongly suppressed back-scattering are formed in this model in the direction of the lines. The system reveals L\'evy-flight transport in stripe direction such that the corresponding conductivity increases as the square root of the system length. Thus, adding this type of disorder to clean graphene near the Dirac point strongly enhances the conductivity, which is in stark contrast with a fully random distribution of on-site impurities which leads to Anderson localization. The effect is demonstrated both by numerical simulations using the Kwant code and by an analytical theory based on the self-consistent $T$-matrix approximation.
\end{abstract}

\maketitle

In recent years a remarkable progress towards controllable deposition of adatoms such as hydrogen and fluorine on graphene has been achieved \cite{Elias09, Robinson10, Nair10, Withers10,Hong,Stabile}. This development has been motivated in part by an attempt to transform graphene into a two-dimensional semiconductor with a controllable band-gap by producing a large density of adatoms \cite{Sluiter03}. If the impurity concentration is sufficiently small, adatoms (like H, F and Cl) can be accurately modeled by effective on-site potentials in the standard tight-binding Hamiltonian of graphene \cite{Lichtenstein}.

Quantum transport properties of graphene near the Dirac point with various types of disorder have attracted a great deal of attention in recent years. One of the remarkable experimental observations was that of Dirac-point conductivity (``minimal conductivity'') of disordered graphene that is of the order of the quantum value $e^2/h$ but remains temperature-independent down to very low temperatures (30 mK) instead of showing the expected suppression due to Anderson localization. Theoretical works demonstrated that graphene with particular kinds of disorder realize a variety of universality classes (distinguished by symmetries and topologies) and may thus avoid Anderson localization by showing quantum criticality with a scale-independent conductivity $\sim e^2/h$ or antilocalization behavior (with logarithmically increasing conductivity) \cite{suzuura02, Aleiner06, Ostrovsky06, Bardarson07, Ostrovsky10, roche13, Gattenloehner14, laissardiere14, ferreira15}.

In this paper, we demonstrate another type of unconventional transport regime in graphene with disorder formed by adatoms. We show that in the case of anisotropic disorder, with all adatoms located within a set of parallel stripes, the transport along the stripes becomes superdiffusive. This kind of stochastic process known as L\'evy flight \cite{levy} is characterized by a heavy-tailed (power-law) distribution of lengths of elementary steps of ballistic propagation between the consecutive scattering events. In one-dimensional (1D) geometry, the power-law random banded matrix model \cite{mirlin96} represents a quantum transport problem with superdiffusive classical dynamics. Despite its 1D character, this model undergoes a localization-delocalization transition with changing the fat-tail exponent. Normally, this type of behavior is not encountered in disordered systems with finite-range scatterers: one finds a conventional classical diffusion supplemented by quantum localization effects. A notable exception is provided by a problem of a quasi-1D system with surface disorder \cite{leadbeater98} where one finds L\'evy-flight behavior on the quasiclassical level. It yields, however, only a logarithmic enhancement of the quasiclassical diffusion constant and thus does not essentially affect the Anderson localization characteristic for the 1D geometry. As we show in this paper, a striped disorder in graphene [i.e., in a two-dimensional (2D) geometry] leads to a much more striking modification of transport properties, suppressing localization and inducing a square-root increase of conductivity with the system size.

\begin{figure}
\centerline{\includegraphics[width=0.9\columnwidth]{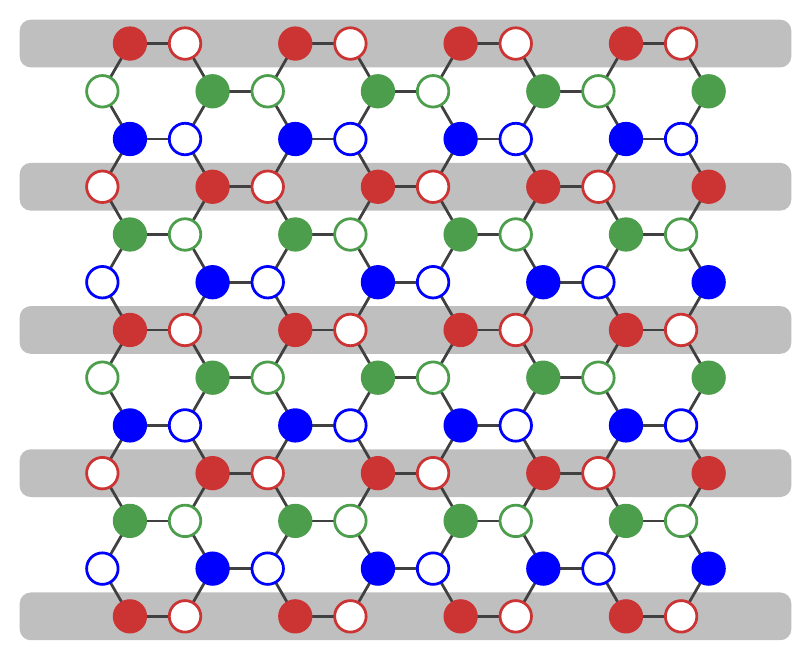}}
\caption{(Color online) The anisotropic disorder model is obtained by randomly placing adatoms on
sites indicated by gray horizontal stripes, with equal probability for sites on sublattices A and
B (filled and empty red circles, respectively).
\vspace*{-0.3cm}}
\label{fig:model}
\end{figure}

The anisotropic impurity distribution is obtained by placing adatoms on every third line drawn along carbon bonds as illustrated in Fig.~\ref{fig:model}. Otherwise, adatoms take random positions and the probabilities to find an adatom on A (filled circles) or B (empty circles) sub-lattice on the line are equal.   In the terminology of Ref.~\cite{Schelter11}, the anisotropic distribution corresponds to restricting adatom positions to the sites of a certain color (red one in Fig.~\ref{fig:model}). The site color refers to the Bloch phase of the zero-energy wave function of the corresponding tight-binding model. It is always possible to choose the gauge such that the Bloch phase takes values $\pm2\pi/3$ or $0$. The relative Bloch phase between the sites of the same color is zero independent of gauge. Each adatom is modeled by an on-site potential $V_0$ which defines the corresponding length scale $\ell_a= V_0 \tilde{a}/t$ with $\tilde{a}=2\sqrt{3} a$, where $a\approx 1.42$\,{\AA} is the length of the carbon-carbon bond and $t\approx 2.7$\,eV is the nearest-neighbor hopping integral in graphene. 

%%%%%%%%%%%%%%%%%%%%%%%%%%%%
%%%% fig:model
%%%%%%%%%%%%%%%%%%%%%%%%%%%%
%%%%%%%%%%%%%%%%%%%%%%%%%%%%%

\begin{figure}
\centerline{\includegraphics[width=\columnwidth]{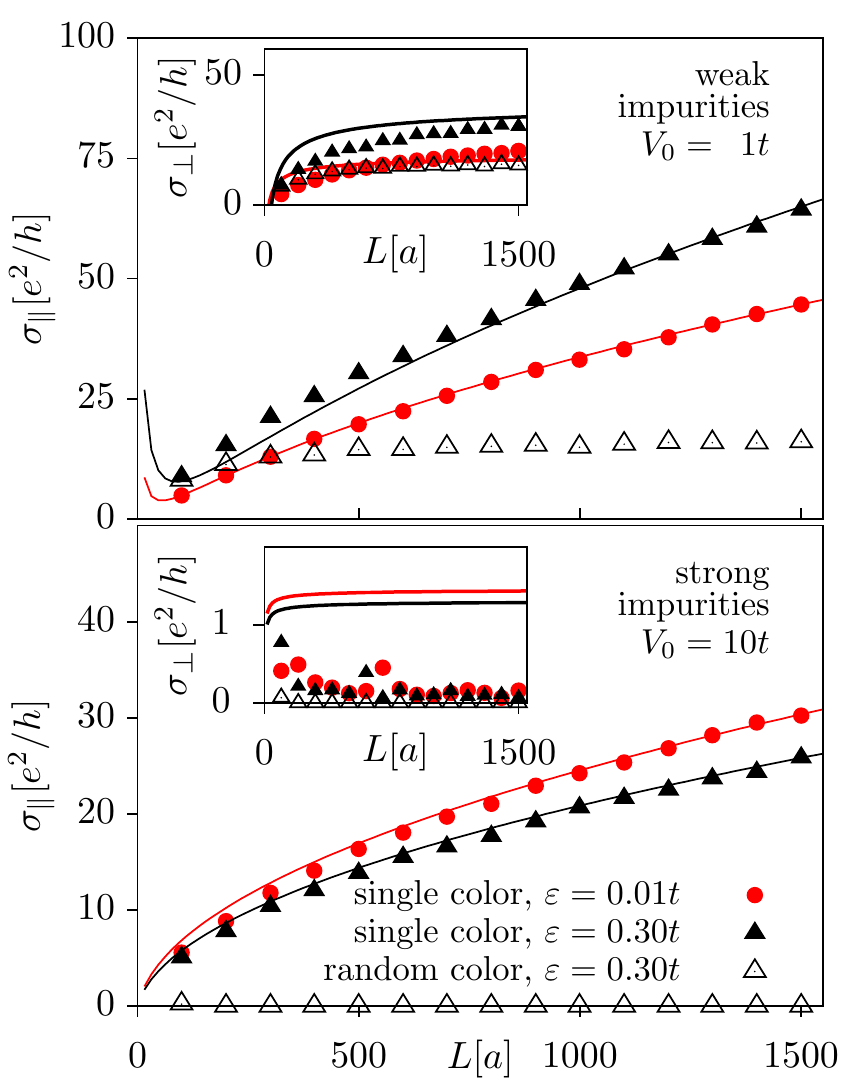}}
\caption{(Color online) Conductivity of graphene with adatom disorder, as evaluated numerically using the Kwant package \cite{kwant}, plotted vs. the system size $L=L_\parallel$. The impurity density is $n_{\rm imp}=0.1/a^2$. Filled symbols correspond to single-color disorder. The conductivity along the stripe direction $\sigma_{\parallel}=\sigma_{xx}$ shows a $\sqrt{L}$ increase, in agreement with the prediction of the Drude theory of Eqs.~(\ref{sigma})--(\ref{Pres}) (solid lines), both for comparatively weak (top panel) and strong (bottom panel) impurities. Insets display the conductivity in the transverse direction, $\sigma_{\perp}=\sigma_{yy}$, which shows a saturation, also in agreement with the Drude theory. The saturation value in the case of strong disorder is, however, much lower than predicted by the SCTMA. For comparison, the corresponding data for impurities distributed randomly over sites of all colors are shown by empty triangles. These data show the conventional behavior---diffusion (top panel) and strong Anderson localization (bottom panel).
\vspace*{-0.3cm}}
 \label{fig:kwant1}
\end{figure}

We employ the Kwant software \cite{kwant} to calculate the averaged conductance of a disordered sample with the dimensions $L_\perp\times L_\parallel$, where $L_\parallel$ stands for the sample length in stripe direction. Two highly doped ballistic graphene leads are attached to the opposite sides of the sample to obtain the conductance in stripe direction $G_\parallel$ and in the direction perpendicular to the stripes, $G_\perp$. The corresponding two-terminal conductivity is obtained from the relations $\sigma_\parallel = G_\parallel L_\parallel/L_\perp$ and $\sigma_\perp= G_\perp L_\perp/L_\parallel$. From the symmetry point of view, the model belongs to Wigner-Dyson orthogonal class (class AI \cite{Altland97}). One can thus expect, in view of the 2D character of the system, a conventional Anderson-localization behavior of the conductivity, i.e., its decrease (exponential in the strong-localization regime) with the system length in the transport direction. This behavior is indeed observed when adatoms are placed randomly on all lattice sites, independently of their colors \cite{Gattenloehner14}. The corresponding data are shown in Fig.~\ref{fig:kwant1} by empty triangles. In the upper panel, where the results for weak impurities, are displayed, the dimensionless Drude conductivity $\sigma/(e^2/h)$ is large, so that the localization length (which increases exponentially with $\sigma$ in 2D) is much larger than our system sizes. Thus, the data show an essentially constant conductivity (diffusive regime). In the lower panel, where the data for stronger impurities are presented, the Drude conductivity is below $e^2/h$, and we observe a strong Anderson localization, as expected \cite{sup}.

%%%%%%%%%%%%%%%%%%%%%%%%%%%%
%%%% fig:kwant1
%%%%%%%%%%%%%%%%%%%%%%%%%%%%
%%%%%%%%%%%%%%%%%%%%%%%%%%%%%

Remarkably, restriction of adatoms to sites of a single color---which is the subject of the present work---turns out to lead to a totally different behavior. Our central result is illustrated in Fig.~\ref{fig:kwant1} by full symbols, with circles corresponding to an energy very close to the Dirac point, $\varepsilon/t = 0.01$, and triangles corresponding to a higher energy, $\varepsilon/t = 0.3$. In the main panels of Fig.~\ref{fig:kwant1}, the conductivity $\sigma_\parallel$ along the stripe direction is presented as a function of the length $L_\parallel$. The upper panel corresponds to the case of weak impurities. For not too large system size, $L_\parallel = 100\;a$, the conductivity is in this case close to that of clean graphene. Indeed, the first full triangle almost coincide with the corresponding empty triangle (random-color disorder), indicating that the type of disorder does not matter. The situation changes, however, with increasing $L_\parallel$. While the system with random-color disorder shows a conventional diffusive regime (discussed above), the conductivity of a system with single-color disorder keeps increasing as $\sigma_\parallel \propto \sqrt{L_\parallel}$.

The difference between the effects of random-color and single-color disorder becomes even more dramatic for the case of stronger scatterers (lower panel). While the random-color model shows in this case a strong suppression due to Anderson localization (as discussed above), the conductivity of the single-color problem keeps showing a square-root increase, $\sigma_\parallel \propto \sqrt{L_\parallel}$. This anomalous behavior is observed for any value of energy, both at the Dirac point and arbitrarily far from it, and is thus a generic property of the lattice model (i.e., it does not depend on Dirac linearization of the Hamiltonian, which is only valid for small energies) \cite{note-vacancies}.

The phenomenon observed can be understood already at the level of the Dirac Hamiltonian 
\be
\label{model0}
H_D=\hbar v\,\bb{p\sigma} + V(\bb{r}),
\e
that applies for $\ep\ll t$. The model (\ref{model0}) assumes that the concentration of adatoms is small, i.e., the average distance between adatoms is much larger than the lattice spacing. Here we use the valley-symmetric representation where $\hbar v=3 ta/2$ and $\bb{\sigma}=(\sigma_x,\sigma_y)$ is the vector of Pauli matrices acting in the sublattice space. The velocity $v$ is set to unity below. The term  $V(\bb{r}) = \sum_i \lt[V_A \delta(\bb{r}-\bb{r}^A_i) + V_B \delta(\bb{r}-\bb{r}^B_i)\rt]$ represents the disorder potential due to on-site adatoms \cite{Basko08, Titov10, Ostrovsky10, Schelter11, Gattenloehner14},  where $\bb{r}^{A(B)}_i$ stand for random adatom positions on the $A(B)$ sublattice. For anisotropic disorder with stripes along x-direction, one finds \cite{Ostrovsky10, Schelter11, Gattenloehner14}, $V_{A(B)}=\ell_a\lt(1+\sigma_x\tau_x\mp \sigma_y\tau_y\pm \sigma_z\tau_z\rt)/4$, where $\ell_a= V_0 \tilde{a}/t = 2\sqrt{3}V_0 at$ and the Pauli matrices $\tau_{x,y,z}$ act in the valley space. The averaging over disorder is, then, performed within the self-consistent T-matrix approximation (SCTMA).  

A specific feature of the anisotropic disorder model constructed above is that the disorder potential can be made diagonal in both sublattice and valley spaces by a global rotation, such that $U\h V_{A(B)} U = \ell_a P_{A(B)}$, where $U=(\tau_z+\sigma_x\tau_x)/\sqrt{2}$ and $P_{A(B)} = (1\pm\sigma_z) (1+\tau_z)/4$. The rotation clearly commutes with the operators of coordinate and momentum. It is, therefore, natural to study the problem in the rotated basis $H_U=U\h H_D U$, 
\be
\label{ham}
H_U= H_{\bb{p}}+ V_U(\bb{r}),\quad H_{\bb{p}}=\sigma_x p_x +\sigma_z \tau_y p_y,
\e
where $V_U(\bb{r}) = \sum_i \lt[P_A \delta(\bb{r}-\bb{r}^A_i) + P_B \delta(\bb{r}-\bb{r}^B_i)\rt]$. The transformed disorder potential, $V_U(\bb{r})$, is present only in a single valley of the rotated model, Eq.~(\ref{ham}). The valley mixing is absent for $p_y=0$, hence the states with small $p_y$ are very weakly affected by disorder. This property is responsible for the quasi-ballistic L\'evi-flight transport in the stripe direction.

In diffusive (Drude) approximation the conductivity is given by
\be
\label{sigma}
\sigma_{aa}=2\frac{2e^2}{h}\int\frac{\mathrm d^2 p}{(2\pi)^2}
                            \tr\lt[j_a\im G^R_{\bb{p}}\rt]^2=\frac{4e^2}{\pi h}\Pi_{aa},
\e
where $G^R_{\bb{p}}=\lt[\ep-H_{\bb{p}}-\Sigma^R\rt]^{-1}$ stands for the retarded disorder-averaged Green's function in the SCTMA. The self-energy takes the form $\Sigma^R=2 s (P_A+P_B)$, where the complex parameter $s=s_0-i\Gamma$ satisfies a non-linear self-consistency equation  (see Supplemental Material \cite{sup}). The exact form of this equation is not important for our analysis below. In full analogy with the case of on-site impurities randomly distributed over sites of all colors \cite{Ostrovsky06}, the vertex corrections to the current operators $j_x=\sigma_x$, $j_y=\sigma_z\tau_y$ are absent.

Integration over momentum in Eq.~\eqref{sigma} develops a very anisotropic singularity for $|(p_x\pm \ep)\ep_0| \propto p_y^2 \to 0$, where $\ep_0=\ep-s_0$. This singularity is non-integrable for the case of $\sigma_{\parallel}=\sigma_{xx}$ in the thermodynamic limit and is regularized by taking into account a finite size of the system, so that $p_y>1/L_\perp$ and $|p_x\pm \ep| > 1/L_\parallel$. In fact, due to the anisotropic character of the singularity, the length $L=L_\parallel$ turns out to be the only relevant regularization parameter, while $L_\perp$ can be regarded infinite. Upon this regularization, the polarization operators $\Pi_{aa}$ can be written as functions of the two dimensionless quantities $\zeta=\ep_0/\Gamma$ and $\delta=\kappa/\Gamma L$, where $\kappa$ is a positive real number that will be used as a fitting parameter. The results are expressed in the form of integrals over $p=(\pm p_x-\ep)/\Gamma$ and $q=p_y/\Gamma$ \cite{sup},
\beml
\beq
\Pi_{xx}&=&\frac{2}{\pi} \int_{-\infty}^\infty\!\!\mathrm dp \int_{-\infty}^\infty\!\! \mathrm dq
                    \frac{(p^2-q^2)^2\; \theta(|p|-\delta)}{\lt| p^2+q^2+2p(\zeta+i)\rt|^4},\qquad\\
\Pi_{yy}&=&\frac{2}{\pi} \int_{-\infty}^\infty \!\!\mathrm dp\int_{-\infty}^\infty \!\! \mathrm dq
                    \frac{4p^2q^2\;\theta(|p|-\delta)}{\lt| p^2+q^2+2p(\zeta+i)\rt|^4},
\eq
\eml
where $\theta(x)$ is the Heaviside theta function. The integrations over $q$ can be performed analytically. The subsequent integration over $p$ can be easily carried out in the limit $\delta\to 0$ (which corresponds to the limit of large system size $L$), with the result
\beml
\label{Pres}
\beq
\Pi_{xx}&=&\frac{A}{2\sqrt{\delta}} - B + \frac{7}{8}C \sqrt{\delta}+\mathcal{O}(\delta^{3/2}),\\
\Pi_{yy}&=&B - C \sqrt{\delta} +\mathcal{O}(\delta^{3/2}).
\eq
\eml
The coefficients $A$, $B$, and $C$ depend on the parameter $\zeta=\ep_0/\Gamma$ as follows:
\beml
\label{Coeff}
\beq
A(\zeta)&=&\re\lt[(1-2i\zeta)\sqrt{1+i\zeta}\rt],\\
B(\zeta)&=&1+\zeta \arctan \zeta,\\
C(\zeta)&=&\re\lt[(1+2i\zeta)/\sqrt{1+i\zeta}\rt].
\eq
\eml

At the Dirac point ($\zeta=0$) we simply find $A=B=C=1$, hence
\be
\sigma_{\parallel}= \frac{4e^2}{\pi h} \lt(\frac{1}{2}\sqrt{\Gamma L/\kappa}-1\rt), \quad
\sigma_{\perp}=\frac{4e^2}{\pi h},\quad \ep_0=0,
\e
where we disregard terms that vanish in the limit $L\to \infty$. Sufficiently far from the Dirac point ($\zeta\gg 1$), we obtain from Eqs.~(\ref{Pres}), (\ref{Coeff}) $\Pi_{xx}/\zeta=u-\pi/2+7/8u+\mathcal{O}(u^{-2})$ and $\Pi_{yy}/\zeta=\pi/2-1/u+\mathcal{O}(u^{-2})$, where $u=\sqrt{\ep_0 L/2\kappa}$. Thus, for $\ep_0\gg\{\Gamma,\kappa/L\}$, we have
\be
\sigma_{\parallel} = \frac{4e^2}{\pi h} \frac{\ep_0}{\Gamma}
                       \lt(\sqrt{\frac{\ep_0 L}{2\kappa}}-\frac{\pi}{2}\rt), \quad
\sigma_{\perp} = \frac{2e^2}{h} \frac{\ep_0}{\Gamma}.
\e
Thus, the Drude analysis in combination with the SCTMA predicts that the conductivity $\sigma_\parallel$ increases as a square root of the system size irrespective of energy. This conclusion is unaffected by quantum interference effects which lead to a weak localization correction that is much smaller than the Drude conductivity \cite{sup}. The analytical results are in agreement with numerical simulation of Fig.~\ref{fig:kwant1}. The results of Eqs.~(\ref{sigma})--(\ref{Pres}) are shown in Fig.~\ref{fig:kwant1} with solid lines where $\Gamma$, $\zeta$, and $\kappa$ take the values given in Table~\ref{tab:fit}. For the transverse conductivity $\sigma_\perp$ the Drude + SCTMA calculation yields an $L$-independent result at large $L$. This prediction is also in qualitative agreement with numerical data in the insets of Fig.~\ref{fig:kwant1}. However, for strong impurities (bottom panel), the saturation values for $\sigma_\perp$ appear to be an order of magnitude smaller than the SCTMA predictions. In fact, since the SCTMA yields in this case values of order $e^2/h$, this calculation is not expected to be parametrically controllable, so that the exact result is expected to deviate by a numerical factor of order unity \cite{sup}. It is interesting that this factor turns out to be so significant.

%%%%%%%%%%%%%%%%
% Table tab:fit
%%%%%%%%%%%%%%%%
\begingroup
\begin{table}
\begin{center}
\tabcolsep=0.45em
\begin{tabular}{l|c|c|c|c}
\hline\hline\\[-6pt]
& \multicolumn{2}{c|}{$V_0/t=1$}& \multicolumn{2}{c}{$V_0/t=10$} \\[3pt]
& $\ep/t =0.01$  & $\ep/t =0.3$ & $\ep/t =0.01$  & $\ep/t =0.3$ \\[3pt]
\hline\\[-6pt]
$\Gamma/t$ & $ 0.0134$ & $0.0097$ & $0.522$ & $0.335$ \\[3pt]
$\zeta$ & $9.98$ & $20.3$ & $0.402$ & $0.192$ \\[3pt]
$\kappa$ & $2.8$ & $6.6$  & $0.29$  & $0.2$ \\[3pt]
\hline\hline
\end{tabular}
\end{center}
\vspace*{-10pt}
\caption{\label{tab:fit} Values of the parameters $\Gamma$, $\zeta$, and $\kappa$ used in Fig.~\ref{fig:kwant1} to plot the Drude theory results (solid lines). The calculation of $\Gamma$ and $\zeta$ is described in the Supplemental Material \cite{sup}; $\kappa$ is a fitting parameter.}
\end{table}
\endgroup
%%%%%%%%%%%%%%%%%

\begin{figure}
\centerline{\includegraphics[width=\columnwidth]{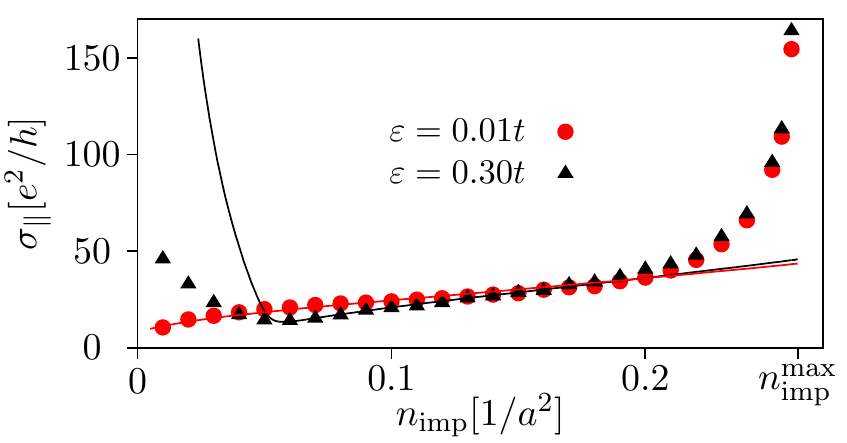}}
\caption{(Color online) Conductivity along stripe direction, $\sigma_\parallel$, calculated using Kwant \cite{kwant} for graphene with single-color impurities at $\ep=0.3 t$ (triangles) and $\ep=0.01 t$ (circles) as a function of impurity concentration. The impurity strength is $V_0=10 t$; the system size is fixed to $L=999.5\,a$. The maximal impurity concentration $n_{\rm imp}^{\rm max}\simeq 0.26/a^2$ corresponds to impurities occupying all allowed sites. The solid lines show the results according to Eqs.~(\ref{sigma})--(\ref{Pres}) with the fitting parameter $\kappa$ as given in Table~\ref{tab:fit} and $\Gamma, \zeta$ as calculated in the Supplemental Material
\cite{sup}.
\vspace*{-0.3cm}}
 \label{fig:kwant2}
\end{figure}

For the sake of completeness, we illustrate in Fig.~\ref{fig:kwant2} the dependence of the conductivity on the concentration of impurities. In a finite sample there exists an ultimate impurity concentration $n_{\rm imp}^{\rm max} \simeq 0.26/a^2$ that corresponds to placing adatoms on all available sites of the given color. For the case when all adatoms have the same potential $V_0$ (as in our model), the resulting system will be strictly periodic, which implies the ballistic character of transport, i.e., an infinite conductivity in the thermodynamic limit. While this behavior is obviously beyond the scope of the effective model, Eq.~(\ref{model0}), it is clearly seen numerically in Fig.~\ref{fig:kwant2}: the conductivity shows a sharp increase when the concentration approaches $n_{\rm imp}^{\rm max}$.

%%%%%%%%%%%%%%%%%%%%%%%%%%%%
%%%% fig:kwant2
%%%%%%%%%%%%%%%%%%%%%%%%%%%%
%%%%%%%%%%%%%%%%%%%%%%%%%%%%%

Finally, it is worth mentioning that the observed L\'evy-flight transport is in fact not specific to graphene but can be found also in other tight-binding models. A simple example is a square lattice with impurities randomly distributed over sites of every second horizontal row. We have checked that the SCTMA predicts a $\sqrt{L}$ increase of $\sigma_\parallel$ also in this case and verified this behavior by numerical simulations \cite{sup}. On the other hand, graphene is a paradigmatic realization of a truly 2D material, and engineering special types of disorder on graphene in a controllable way appears to be within experimental feasibility. This explains our focus on the graphene model in the present paper.

A possible experimental realization of the anisotropic disorder can utilize the macroscopic self-orientation of graphene on hexagonal boron nitride reported recently in Ref.~\cite{1D-16}. The interplay of van der Waals and elastic forces in such structures has been shown to lead to a spontaneous quasi-one-dimensional wrinkling (uniaxial straining) of graphene, similar to Moir{\'e} patterning (see, e.g., Ref.~\cite{Moire} and references therein). Such a self-alignment of 2D crystals naturally produces a one-dimensional potential for adatoms or molecules that might be favorable for creating striped disorder.

In conclusion, we have studied the conductivity of graphene with on-site impurities (adatoms) randomly distributed over striped locations. We have shown that the system reveals L\'evy-flight transport in stripe direction, so that the conductivity $\sigma_\parallel$ increases as the square root of the system length. This behavior is in stark contrast with the Anderson localization observed for a fully random distribution of on-site impurities. 
We hope that this work will pave a way to a long-sought experimental realization of L\'evy flights in electronic quantum transport.

We acknowledge useful discussions with R.~Claessen, R.~Danneau, R.~Krupke, V.~Meded, K.\,S.~Novoselov, and M.~Ruben on possible experimental realizations of the model. We are especially thankful to J.~Schelter who participated in the first stage of this work. The work was supported by the Dutch Science Foundation NWO/FOM 13PR3118, by the Russian Science Foundation under Grant No.\ 14-42-00044 (I.\,V.\,G., P.\,M.\,O., and A.\,D.\,M.), by SPP 1459 of the Deutsche Forschungsgemeinschaft, and by the EU Network FP7-PEOPLE-2013-IRSES under Grant No. 612624 ``InterNoM''.

%================Supplementary===============
\supplementarystart

\centerline{\bfseries\large ONLINE SUPPLEMENTAL MATERIAL}
\vspace{6pt}
\centerline{\bfseries\large L\'evy flights due to anisotropic disorder in graphene}
\vspace{6pt}
\centerline{S.~Gattenl{\"o}hner, I.\,V.~Gornyi, P.\,M.~Ostrovsky, B.~Trauzettel, A.\,D.~Mirlin and
            M.~Titov}

\begin{quote}
In this Supplemental Material, we (i) provide details of the self-consistent T-matrix approximation
which was used to relate the real and imaginary part of the self-energy to the
microscopic parameters of the model (the Fermi energy $\ep$, the impurity strength $V_0$,
and the impurity concentration $\nimp$), (ii) describe how the numerical simulations were
performed, (iii) present results indicating that superdiffusive transport can also be
realized on a square lattice, and (iv) discuss interference corrections to the SCTMA results.
\end{quote}

\section{Disorder-averaged Green's function in the self-consistent T-matrix approximation}

\begin{figure}[htpb]
    \centering
    \includegraphics[width=0.9\columnwidth]{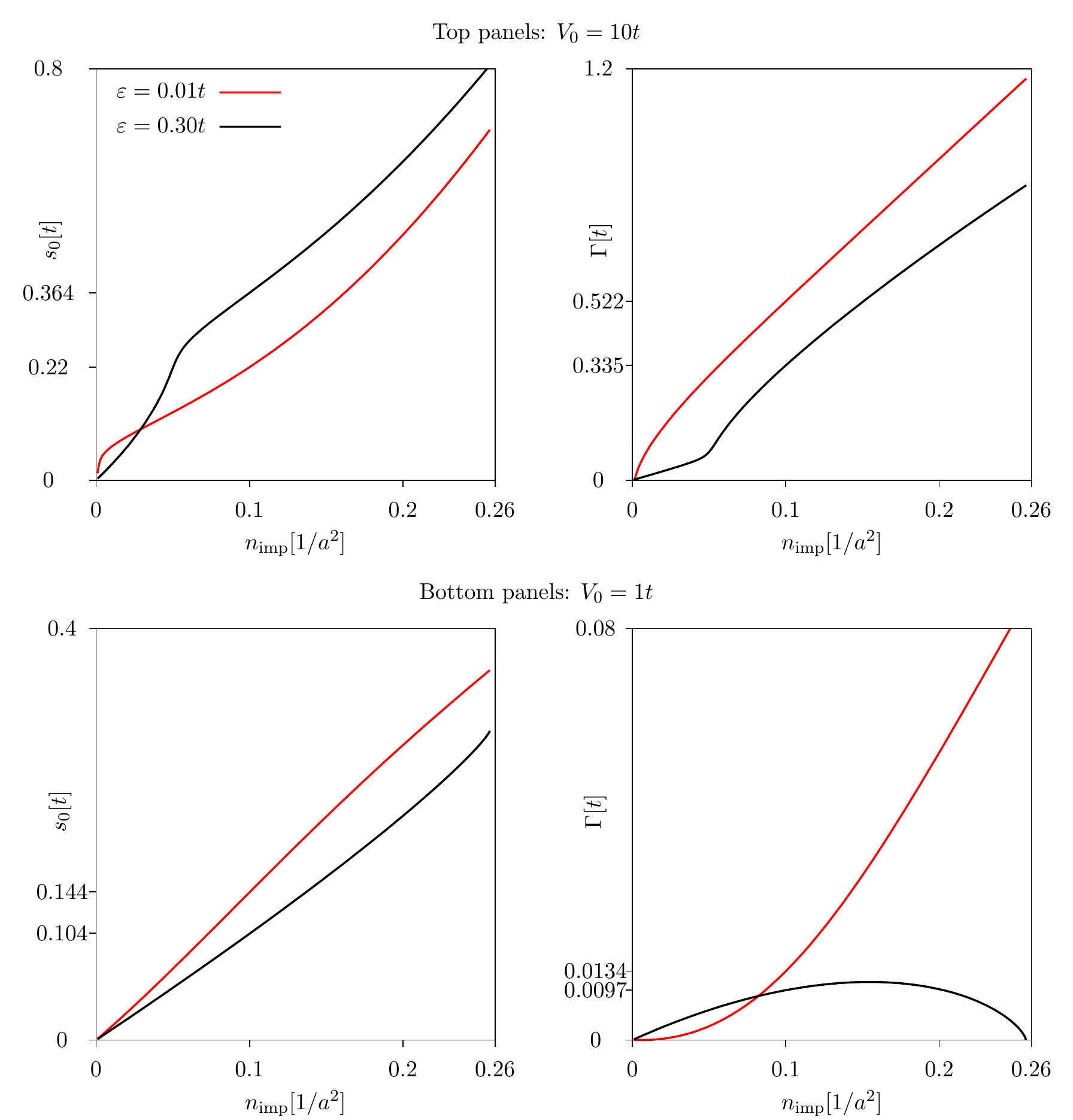}
    \caption{Numerical solution of the self-consistency equation \eqref{s} for the scalar self-energy
    $s$ as a function of the impurity density $\nimp$. The data was computed numerically
    using Newton's method as described in Eq.~\eqref{eq:newton} for the two values of the impurity strength
    $V_0$ and the two values of the energy $\varepsilon$ used in Figures 2 and 3 of the main text.}
    \label{fig:s}
\end{figure}

In this Section of the Supplemental Material, we outline the calculation of the disorder-averaged
Green's function that we use in Eq.~\eqref{sigma} to evaluate the conductivities $\sigma_\parallel$
and $\sigma_\perp$. We employ the self-consistent T-matrix approximation (SCTMA) that neglects
contributions from diagrams with intersecting impurity lines. The implementation of
this approximation scheme for the model (\ref{model0}) is fully analogous to that
described in Ref.~\onlinecite{Ostrovsky06sup} where graphene with ad-atoms of \emph{random} color is
studied. For the \emph{single-color} case, it is convenient to perform these calculations in the rotated
basis of Eq.~(\ref{ham}), hence one readily obtains the expression
\begin{equation}
\label{Sigma}
\hat{\Sigma} = \nimp \left\langle T\right\rangle =
               \frac12 \nimp \left(
                                \ell_a P_A \frac{1}{1-\ell_a g P_A} +
                                \ell_a P_B \frac{1}{1-\ell_a g P_B}
                             \right)
\end{equation}
for the self-energy $\hat\Sigma$, where $\nimp$ is the impurity concentration and $g$ denotes the
averaged Green's function in SCTMA at coinciding real-space arguments,
\begin{equation}
 g = \int\frac{\mathrm d^2p}{(2\pi)^2}\frac{1}{\ep-H_{\bb{p}}-\hat{\Sigma}},
 \quad \text{with } H_{\bb{p}}=\sigma_x p_x +\sigma_z \tau_y p_y.
 \label{g}
\end{equation}
As a consequence of Eq.~(\ref{Sigma}), the (retarded) self-energy is of the form
\begin{equation}
\label{Sigma_simplified}
    \hat{\Sigma} =2sP,
    \quad \text{with } P = P_A+P_B,
    \quad P_A=\left(\begin{smallmatrix}1&0&0&0\\0&0&0&0\\0&0&0&0\\0&0&0&0\end{smallmatrix}\right),
    \quad \text{and }P_B=\left(\begin{smallmatrix}0&0&0&0\\0&1&0&0\\0&0&0&0\\0&0&0&0\end{smallmatrix}\right),
\end{equation}
where we also introduced the scalar self-energy $s=s_0-i\Gamma$. Here $s_0$ and $\Gamma > 0$ are real
numbers. The scalar self-energy satisfies the equation
\begin{equation} \label{s}
s = \frac{\nimp\ell_a}{4(1-g_0\ell_a)},
\end{equation}
where $g_0$ denotes the $11$- and $22$-component of the matrix Green function $g$ with coinciding spatial arguments, Eq.~(\ref{g}):
\begin{subequations}
\begin{align}
\label{firstline}
g_0=g_{11}=g_{22}&=
     \frac{1}{2}\int \frac{\mathrm d^2p}{(2\pi)^2}
     \left(\frac{\ep+p_x}{\ep^2-p_x^2-p_y^2-2s(\ep+p_x)}
     +\frac{\ep-p_x}{\ep^2-p_x^2-p_y^2-2s(\ep-p_x)} \right)\\
  &=\int\frac{\mathrm d^2p}{(2\pi)^2}\frac{\ep-p_x}{\ep^2-p_x^2-p_y^2-2s(\ep-p_x)}=
    \int\!\!\frac{\mathrm d^2p}{(2\pi)^2}\frac{\varepsilon-s-p_x}{(\varepsilon-s)^2
          -p_x^2-p_y^2}.
\label{secondline}
\end{align}
\end{subequations}
The first expression in Eq.~(\ref{secondline})  is obtained by symmetrizing the integrand with the help of the transformation $p_x\to -p_x$ in the first fraction in Eq.~(\ref{firstline}). In the last expression in Eq.~(\ref{secondline}) we have also used the shift $p_x\rightarrow p_x +s$. The resulting integral features an
ultraviolet divergence that is regularized by introducing a cut-off in $p_x$,
\begin{equation}
g_0 =
    \int\limits_{-\frac{\Delta}{2}}^{\frac{\Delta}{2}}\!\frac{\mathrm dp_x}{2\pi}
    \int\limits_{-\infty}^{\infty}\!\frac{\mathrm dp_y}{2\pi}
            \frac{\varepsilon-s-p_x}{(\varepsilon-s)^2-p_x^2-p_y^2}.
\end{equation}
This integral can be evaluated analytically, yielding
\begin{equation}
\label{g0}
\begin{split}
g_0 &=
-\frac{1}{4\pi}
  \left[\rule{0pt}{1cm}\Delta\left(
    \sqrt{
      \left(\frac12-\frac{\varepsilon-2s}{\Delta}\right)
      \left(\frac12+\frac\varepsilon\Delta\right)
    }
    -\sqrt{\left(\frac12+\frac{\varepsilon-2s}{\Delta}\right)
           \left(\frac12-\frac\varepsilon\Delta\right)}
                             \right)
  \right.\\
  &\quad
    \left. +
    \left(\varepsilon-s\right) \ln\left[
        -\frac{\Delta^2 \left(
                           \sqrt{\frac12-\frac{\varepsilon-2s}{\Delta}}
                           +\sqrt{\frac12+\frac\varepsilon\Delta}
                        \right)^2
                        \left(
                           \sqrt{\frac12+\frac{\varepsilon-2s}{\Delta}}
                           +\sqrt{\frac12-\frac\varepsilon\Delta}
                        \right)^2}
              {4(\varepsilon-s)^2}
                                  \right]
   \rule{0pt}{1cm}\right].
\end{split}
\end{equation}
Equation \eqref{s}, with the result \eqref{g0} for $g_0$ inserted, represents a non-linear equation
for the scalar self-energy $s$ that we will now solve numerically. To that end we rewrite
Eq.~\eqref{s} as $f(s)=0$ with
\begin{equation}
    f(s) = (1 - \ell_ag_0)s - \nimp\ell_a/4
\end{equation}
and apply Newton's method assuming that the iteration
\begin{equation}
    \label{eq:newton}
    s^{(n+1)} = s^{(n)} -\frac{f\left(s^{(n)}\right)}{f'\left(s^{(n)}\right)}
\end{equation}
converges to the solution for a suitably chosen initial value $s^{(0)}$. Fig.~\ref{fig:s} shows
the real and imaginary part of the scalar self-energy as a function of the impurity density $\nimp$
for two values of the energy, $\varepsilon=0.01t$ and $\varepsilon=0.3t$,
and two values of the impurity strength, $\ell_a = 2\sqrt{3}a$ (corresponding to $V_0=t$) and
$\ell_a=20\sqrt{3}a$ (corresponding to $V_0=10t$). The ultraviolet cut-off provided by the lattice constant was fixed at
$\Delta =2.5\hbar v/a$.

%%%%%%%%%%%%%%%%%%%%%%%%%%

\section{Derivation of Eq. (4) of the main text}

In this section, formula \eqref{sigma} of the main text,
\begin{equation}
    \label{sigma_supp}
    \sigma_{aa}=2\frac{2e^2}{h}\int\frac{\mathrm{d}^2p}{(2\pi)^2}
                            \tr\lt[j_a\im G^R_{\bb{p}}\rt]^2=\frac{4e^2}{\pi h}\Pi_{aa},
\end{equation}
for the longitudinal ($a=x$) and the transveral ($a=y$) conductivity is evaluated
and brought into the form of Eq.~(4). We perform this calculation in the
rotated basis of Eq.~\eqref{ham}, such that the current operators
are given by $j_x=\sigma_x$ and $j_y=\sigma_z\tau_y$, and use the result of the
previous section that the averaged retarded Green's function in SCTMA is given by
$G^R_{\boldsymbol{p}}=[\varepsilon-H_{\boldsymbol{p}}-2s(P_A+P_B)]^{-1}$. The latter formula is readily eveluated:
\begin{align}
    \label{GR}
G^R_{\boldsymbol{p}} &=
          \frac{
           \left(\begin{smallmatrix}
            p_x-\varepsilon & -(p_x-\varepsilon) & i p_y & ip_y\\
            -(p_x-\varepsilon) & p_x-\varepsilon & -ip_y & -ip_y\\
            -ip_y&ip_y&-(p_x-\varepsilon)+2(s-\varepsilon)&-(p_x-\varepsilon)+2(s-\varepsilon)\\
            -ip_y & ip_y & -(p_x-\varepsilon)+2(s-\varepsilon) & -(p_x-\varepsilon)+2(s-\varepsilon)
           \end{smallmatrix}\right)}
          {2[(p_x-\varepsilon)(p_x+\varepsilon-2s)+p_y^2]}
          +\frac{
           \left(\begin{smallmatrix}
            -(p_x+\varepsilon) & -(p_x+\varepsilon) & ip_y & -ip_y\\
            -(p_x+\varepsilon) & -(p_x+\varepsilon) & ip_y & -ip_y\\
            -ip_y & -ip_y & (p_x+\varepsilon)+2(s-\varepsilon) & -(p_x+\varepsilon)-2(s-\varepsilon)\\
            ip_y &ip_y & -(p_x+\varepsilon)-2(s-\varepsilon) & (p_x+\varepsilon)+2(s-\varepsilon)
           \end{smallmatrix}\right)}
          {2[(p_x+\varepsilon)(p_x-\varepsilon+2s)+p_y^2]}.
\end{align}
Inserting this expression into Eq.~\eqref{sigma_supp} yields an integral whose integrand is
a sum of fractions of the form
\begin{equation}
\frac{F(p_x,p_y,\varepsilon,s_0,\Gamma)}{(p_x\pm\varepsilon)[p_x\mp(\varepsilon-2s)]+p_y^2}
\end{equation}
(and similar expressions involving $s^\ast$), where $F$ is a certain function. This integral can be
further simplified by applying to each term of the integrand either the transformation
$p_x\rightarrow p_x+\varepsilon$ or the transformation $p_x\rightarrow -p_x-\varepsilon$, thereby
shifting the zero of each denominator to the origin. After doing so, we obtain the expressions
\begin{align}
    \Pi_{xx} &= 8\Gamma^2 \int\frac{\mathrm dp_x\,\mathrm dp_y}{4\pi}\ \frac{(p_x^2-p_y^2)^2}
                                                            {|p_x^2+p_y^2+2p_x(\varepsilon-s)|^4},\\
    \Pi_{yy} &= 8\Gamma^2 \int\frac{\mathrm dp_x\,\mathrm dp_y}{4\pi}\ \frac{4p_x^2p_y^2}
                                                            {|p_x^2+p_y^2+2p_x(\varepsilon-s)|^4}.
\end{align}
Rescaling all momenta by $\Gamma$ and introducing the infrared cutoff provided by the system size $L=L_\parallel$, we obtain Eq.~(4) of the main text.

%To compare these results with those found in the numerical simulations on the tight-binding level,
%one needs to adapt above expressions to account for finite sample sizes. A simple way to do
%this is by introducing a cut-off at small momenta, e.g. by neglecting those momenta with
%$|p_x|<p_{x0}=\kappa/L_\parallel$ and
%$|p_y|<p_{y0}=\tilde\kappa/L_\perp$, where $\kappa$ and $\tilde\kappa$ are dimensionless fitting
%parameters. Due to the anisotropic character of the studied model, the conductivities do not strongly
%depend on $p_{0y}$ and the calculations have been further simplified in the main text
%by setting $\tilde\kappa$ to zero.

\section{Implementation of the numerical simulations}

The numerical results presented in Figures 2 and 3 of the Letter were obtained using the Kwant code
\cite{kwant_sup}. This python package allows for a very convenient definition of tight-binding
systems and provides various tools to perform quantum transport calculations on them. We refer the reader to the Kwant reference paper \cite{kwant_sup} for technical aspects of the package.
In this Section of the Supplemental Material we provide a concise description of the
tight-binding models studied in our work.

%%%%%%%%%%%%%%%%%%%%%%%%%%
\begin{figure}[htbp]
    \centering
    \includegraphics[width=0.97\linewidth]{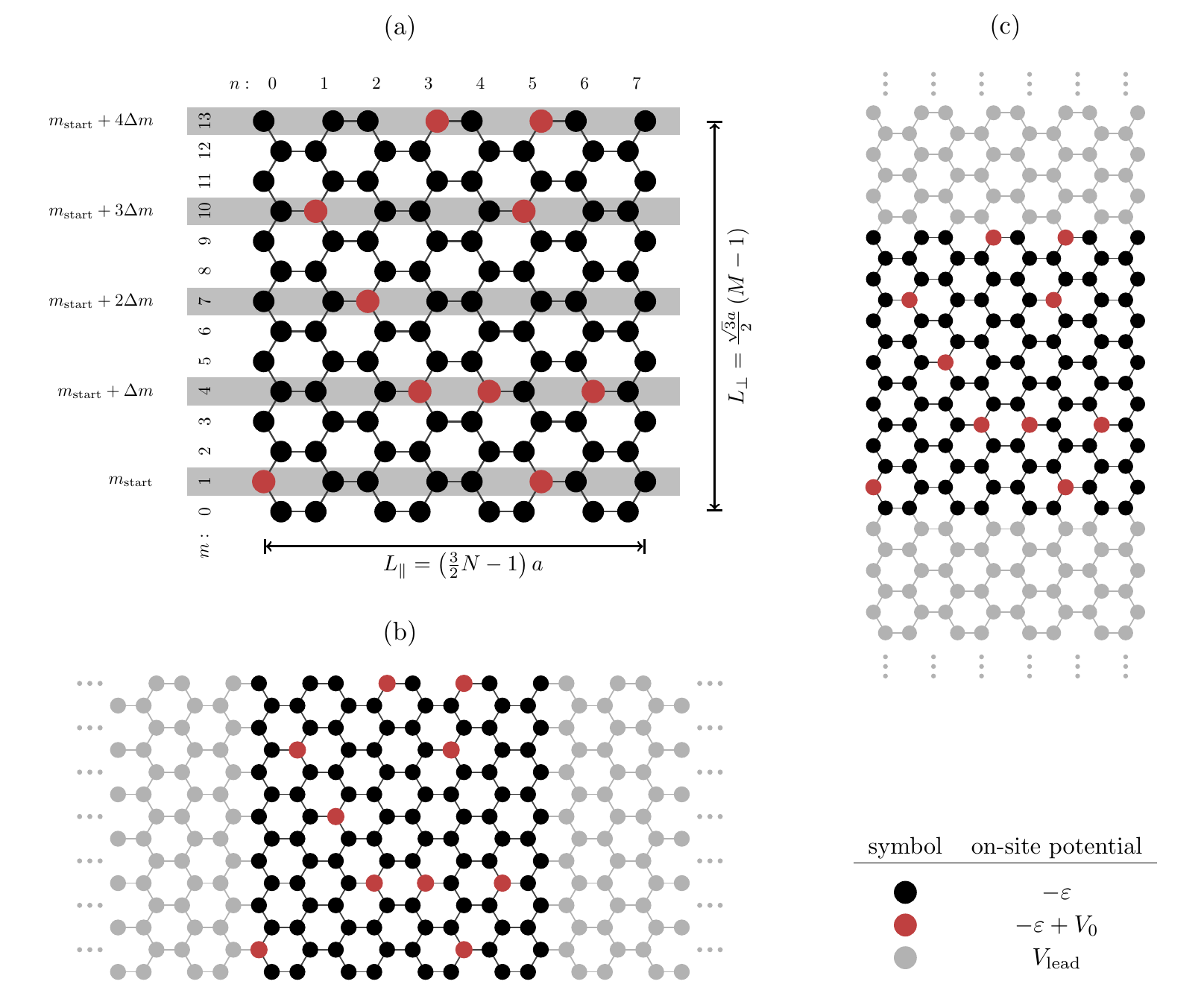}
    \caption{%
(a) Tight-binding model of graphene with impurities (red sites, on-site potential $V_0$) placed randomly within every third horizontal row ($\Delta m=3$).
To compute the conductivities $\sigma_\parallel$ and $\sigma_\perp$,
strongly doped semi-infinite leads are attached to a disordered,
rectangular graphene sample in the direction parallel or perpendicular to the
disordered stripes, respectively, as shown in panels (b) and (c), respectively.}
    \label{fig:kwant_supp}
\end{figure}
%%%%%%%%%%%%%%%%%%%%%%%%%%%%%

We consider rectangular graphene samples with on-site potential disorder that is limited to every
third row. To be specific, we assume samples comprising a definite number of $M$ dimer lines (the
rows, indexed by the letter $m$) and a definite number of $N$ zig-zag lines (the columns, indexed
by the letter $n$), see Figure~\ref{fig:kwant_supp} (a). The size of such a sample is given
by $L_\parallel = (3N/2-1)a$ and $L_\perp = (\sqrt{3} a/2)(M-1)$, where $a$ denotes the nearest
neighbor distance of graphene.\footnote{When the text refers to a sample of a certain length
$L_\parallel$ and a certain aspect ratio, it is assumed that $L_\parallel$ is indeed of the form
$(3N/2-1)a$ and that the number of rows is chosen such that the resulting $L_\parallel/L_\perp$
comes closest to the required aspect ratio (if there is a tie, the larger $M$ is chosen).}

The on-site potentials of such a sample can be arranged in an $M\times N$ matrix
$V_{mn}$ that we set to
\begin{equation}
    V_{mn} =\left\{
        \begin{array}{ll}
            -\varepsilon+V_0,&\text{for $(m,n)$ being a impurity site}\\
            -\varepsilon,& \text{otherwise}.
        \end{array}\right.
\end{equation}

The $N_\mathrm{imp}$ impurity sites are randomly distributed within the allowed rows, where we
call ``allowed'' every third row starting from row $m_\text{start}$. A typical impurity configuration is shown
in Figure~\ref{fig:kwant_supp}~(a) for $m_\text{start}=1$ (for the data shown in Figures 2 and 3 of the Letter,
the choice $m_\text{start}=0$ was used).

To find the conductivities $\sigma_\parallel$ and $\sigma_\perp$, we attach semi-infinite,
strongly doped graphene leads ($V_\mathrm{lead}=-0.3t$, with $t$ being the hopping energy between
neighboring carbon atoms) in the direction parallel or perpendicular to the disordered stripes,
respectively [see Fig.~\ref{fig:kwant_supp} (b) and (c)], and make Kwant calculate the transmissions
$T_{\parallel}$ and $T_\perp$. The conductivities are then given by

\begin{align}
   \sigma_\parallel &= G_\parallel L_\parallel/L_\perp, &\sigma_\perp&= G_\perp L_\perp/L_\parallel,
   &\text{with } \ \ G_{\parallel} &= (2e^2/h) T_{\parallel},  \qquad
   G_{\perp} = (2e^2/h) T_{\perp}.
\end{align}

This calculation is repeated for 50 different disorder configurations to obtain averaged values
for the conductivities.

%%%%%%%%%%%%%%%%%%%%%%%%%%%
\begin{figure}[htpb]
    \centering
    \includegraphics{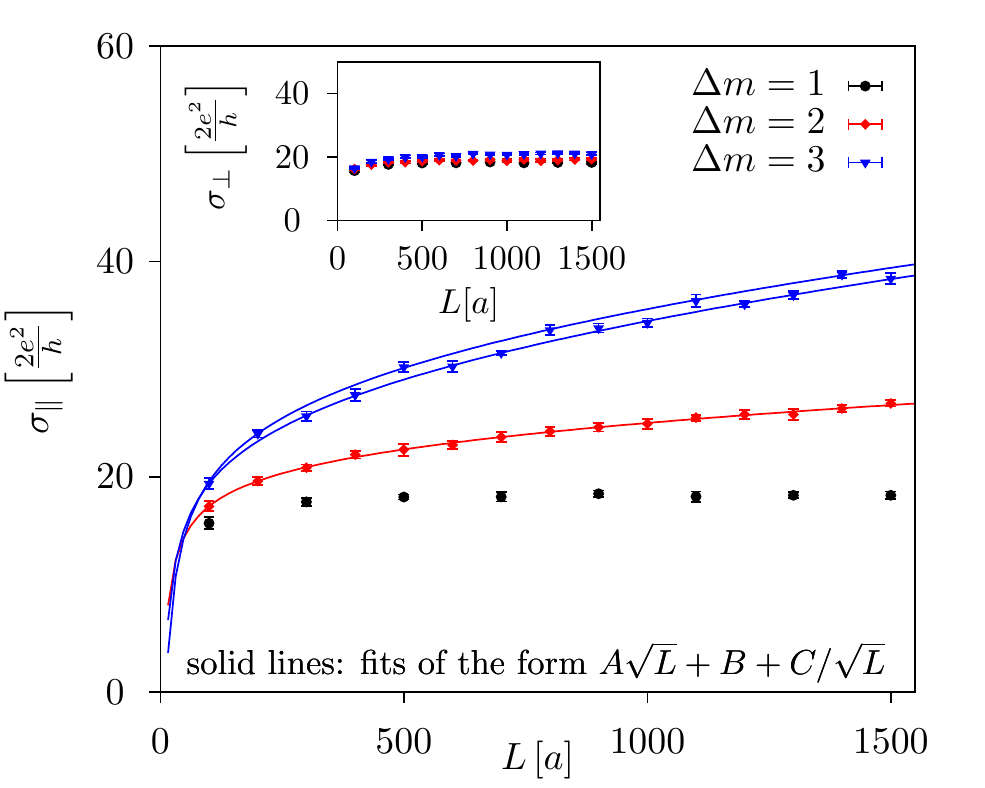}
    \caption{Longitudinal conductivity $\sigma_\parallel$ for the disordered square lattice as a function of the system
		length for fully random on-site impurities ($\Delta m=1$) and
		for on-site impurities that can occur only in every second or third row
		(i.e. $\Delta m=2$ or $\Delta m=3$, respectively). The strength of the impurities
        is given by $V_0=1t$ ($t$ being the hopping energy), the energy is set to
        $\varepsilon=0.3t$, the aspect ratio of the samples is such that $L_\perp=L_\parallel=L$,
        and the impurity density is given by $\nimp=0.1/a^2$ ($a$ being the lattice constant).
        The data was obtained using the Kwant code \cite{kwant_sup}
        and each data point is an average over ten disorder configurations. For $\Delta m=3$ two data sets
        (and, correspondingly, two slightly different fitting lines) are shown, which
        correspond to different boundary scenarios.
 Specifically, samples for which both	terminating rows can contain impurities
		have a longitudinal conductivity that is larger by roughly $2e^2/h$
		than in situations where this cannot happen; for $\Delta m=3$,
		the former is the case whenever $L_\perp = (3 k -1)a$ for some integer $k$. The inset shows
        the conductivity $\sigma_\perp$ in the direction perpendicular to the disordered rows which displays diffusive
        transport independent of the value of $\Delta m$.}
    \label{fig:square}
\end{figure}
%%%%%%%%%%%%%%%%%%%%%%%

\section{Superdiffusive transport on the square lattice}

As pointed out in the end of the main text of the paper, the mechanism of the L\'evy-flight
transport found in our work is in fact not specific to graphene but can be also found in
other tight-binding models. In this Section of the Supplemental Material we present numerical
data obtained using the Kwant code \cite{kwant_sup} for a square lattice where on-site
potential impurities are randomly distributed within every $\Delta m$-th horizontal row.
Figure~\ref{fig:square} shows that for $\Delta m=1$ one observes the usual diffusive transport,
while for $\Delta m=2$ or $\Delta m=3$ one finds superdiffusive transport with a square-root
dependence of the longitudinal conductivity on the system length.

\section{Interference correction: Weak localization}

In a conventional 2D system, Anderson localization sets in when the system size is large enough. To see this, one can calculate the weak localization correction to the conductivity which is negative and diverges logarithmically with the system size $L$. In this section of the Supporting Material, we calculate the weak localization correction for the  problem with the anisotropic disorder of the type studied in the present work. We show that for the longitudinal conductivity $\sigma_\parallel$, the weak-localization correction is much smaller than the SCTMA result obtained in the Letter. As a consequence, the localization does not affect our key conclusion about the $\sqrt{L}$ L\'evy-flight behavior of the longitudinal conductivity. For the transverse conductivity  $\sigma_\perp$, we get a non-singular (at $L\to\infty$) correction of order unity, i.e., of the same order as the SCTMA result. This behavior indicates that the transverse conductivity remains $L$-independent in the large-$L$ limit but with a numerical value renormalized in comparison with SCTMA, in agreement with numerical data.

To calculate the weak-localization correction to the SCTMA results, we follow the general approach
developed by W\"olfle and Bhatt \cite{WB} for anisotropic disordered systems.
According to Eq.~(11) of Ref.~\cite{WB}, the weak localization correction
to the conductivity at frequency $\omega$ can be represented in the following form 
\be 
    \label{WBform}
    \delta\sigma_{\alpha\alpha} = 
    -\frac{4e^2}{h} \s_{\bb{q}}\frac{D_{\alpha\alpha}}{-i\omega+\s_{\beta} D_{\beta\beta} q^2} 
\e 
where $D_{\alpha\alpha}$ stands for the (generally anisotropic) quasiclassical (Drude) diffusion coefficient.
In our model, the anisotropy is so specific that it introduces an anomalous diffusion. The latter
reveals itself in the $q_x$ dependence of $D_{xx}$ such that 
$D_{xx}(q_x) = D_{xx}^{0} (q_x \ell_0)^{-1/2}$, where $\ell_0$ is a length scale representing the ultraviolet (short-distance) cutoff for the diffusion. The diffusion coefficient $D_{yy}$ remains $q$-independent. Using this behavior of the Drude (SCTMA) diffusion coefficients, we readily obtain the conductivity at zero frequency $\omega=0$ by taking the
integral over $\bb{q}$ in Eq.~(\ref{WBform}). The integration over
$q_y$ and $q_x$ is performed separately ($q_y$ is integrated first) in the same way as in the
calculation of the polarization operator. A cut-off at small momenta $q_x$ is introduced.

The resulting expression for the weak-localization correction to the longitudinal conductivity $\sigma_\parallel=\sigma_{xx}$
scales as $L^{1/4}$ for system size $L  \gg l_0$. This is much smaller than the Drude (SCTMA) conductivity that scales as $L^{1/2}$. Thus, the localization effects on the longitudinal conductivity are negligible.

The correction to the transverse conductivity $\sigma_\perp = \sigma_{yy}$ saturates at $L\gg l_0$ at a value of
order of unity,  $\delta\sigma_\perp=\delta \sigma_{yy} \propto - [1-(l_0/L)^{1/4}]$.  This non-singular behavior indicates the absence of the strong Anderson localization also for the transverse direction. On the other hand, the correction is of the same order as the Drude (SCTMA) result. This implies that, while the transverse conductivity remains $L$-independent as in SCTMA, its numerical value is substantially suppressed. This is indeed was is observed in our numerical simulations, see lower panel of Fig.~\ref{fig:kwant1} of the main
text.


\begin{thebibliography}{99}
\bibitem{note-vacancies} It is worth pointing out that, while the observed $\sqrt{L_\parallel}$
increase of the conductivity for graphene with single-color disorder holds equally for weak and
strong impurities, and at any energy, it disappears in the limit of infinitely strong adatom
potential (or, equivalently, vacancies) at the Dirac point. In this limit the conductivity
$\sigma_\parallel$ still does not show Anderson localization but is independent of $L_\parallel$ for
large system sizes \cite{Ostrovsky10}.
\end{thebibliography}
\end{document}